\documentclass[manuscript]{aastex}
\usepackage{emulateapj5}
\usepackage{apjfonts}
\usepackage{epsfig}
\usepackage{lscape}

\journalinfo{The Astrophysical Journal Letters in Press}
\slugcomment{Received 2008 December 19; accepted 2009 April 9}
\shorttitle{EPISODIC RANDOM ACCRETION AND SPINS OF SMBHS}
\shortauthors{WANG ET AL.}

\def\kms{\ifmmode {\rm km~ s^{-1}} \else {\rm km~s^{-1}}\ \fi}
\def\mbh{M_{\bullet}}

\def\mgii{\ifmmode Mg {\sc ii} \else Mg {\sc ii}\ \fi}
\def\oiii{\ifmmode O {\sc iii} \else O {\sc iii}\ \fi}
\def\feii{\ifmmode Fe {\sc ii} \else Fe {\sc ii}\ \fi}
\def\rhobh{\rho_{\bullet}}
\def\rmd{{\rm d}}

\def\sunm{M_{\odot}}

\def\lbol{L_{\rm bol}}

\def\gsim{\mathrel{\rlap{\lower 4pt \hbox{\hskip 1pt $\sim$}}\raise 1pt
\hbox {$>$}}}
\def\lsim{\mathrel{\rlap{\lower 4pt \hbox{\hskip 1pt $\sim$}}\raise 1pt
\hbox {$<$}}}
\def\lax{{$\mathrel{\hbox{\rlap{\hbox{\lower4pt\hbox{$\sim$}}}\hbox{$<$}}}$}}
\def\gax{{$\mathrel{\hbox{\rlap{\hbox{\lower4pt\hbox{$\sim$}}}\hbox{$>$}}}$}}

\begin{document}

\title{Episodic Random Accretion and the Cosmological Evolution of Supermassive Black Hole Spins}

\author{
Jian-Min Wang\altaffilmark{1,6}, 
Chen Hu\altaffilmark{1}, 
Yan-Rong Li\altaffilmark{1}, 
Yan-Mei Chen\altaffilmark{1},
Andrew R. King\altaffilmark{2},
Alessandro Marconi\altaffilmark{3}, 
Luis C. Ho\altaffilmark{4},
Chang-Shuo Yan\altaffilmark{1},
R\"udiger Staubert\altaffilmark{5}, and
Shu Zhang\altaffilmark{1}
}

\altaffiltext{1}
{Key Laboratory for Particle Astrophysics, Institute of High Energy Physics,
Chinese Academy of Sciences, 19B Yuquan Road, Beijing 100049, China}

\altaffiltext{2}
{Theoretical Astrophysics Group, University of Leicester, Leicester LE1 7RH, UK}

\altaffiltext{3}
{Department of Astronomy and Space Science, University of Florence, Largo Enrico Fermi 2, 
50125 Firenze, Italy}

\altaffiltext{4}
{The Observatories of the Carnegie Institution of Washington, 813 Santa Barbara Street, 
Pasadena, CA 91101, USA}

\altaffiltext{5}
{IAAT, Abt. Astronomie, Universit\"at T\"ubingen, Sand 1, 72076 T\"ubingen, Germany}

\altaffiltext{6}
{Theoretical Physics Center for Science Facilities, Chinese Academy of Sciences, 
Beijing 100049, China}

\begin{abstract}
The growth of supermassive black holes (BHs) located at the centers of their host galaxies 
comes mainly from accretion of gas, but how to fuel them remains an outstanding unsolved 
problem in quasar evolution.  This issue can be elucidated by quantifying the radiative 
efficiency parameter ($\eta$) as a function of redshift, which also provides constraints 
on the average spin of the BHs and its possible evolution with time.  We derive a formalism 
to link $\eta$ with the luminosity density, BH mass density, and duty cycle of quasars, 
quantities we can estimate from existing quasar and galaxy survey data. We find that $\eta$ 
has a strong cosmological evolution: at $z\approx 2$, $\eta \approx 0.3$, and by $z\approx 0$ 
it has decreased by an order of magnitude, to $\eta\approx 0.03$.  We interpret this trend 
as evolution in BH spin, and we appeal to episodic, random accretion as the mechanism for 
reducing the spin. The observation that the fraction of radio-loud quasars decreases with 
increasing redshift is inconsistent with the popular notion that BH spin is a critical factor 
for generating strong radio jets. In agreement with previous studies, we show that the derived 
history of BH accretion closely follows the cosmic history of star formation, consistent with 
other evidence that BHs and their host galaxies coevolve.  
\end{abstract}
\keywords{black hole physics --- galaxies: evolution --- quasars: general}

\section{Introduction}
The empirical correlation between black hole (BH) mass and bulge 
luminosity (Kormendy \& Richstone 1995; Magorrian et al.  1998; Marconi \& Hunt 2003) and stellar 
velocity dispersion (Gebhardt et al. 2000; Ferrarese \& Merritt 2000; Tremaine et al. 2002) have 
stimulated the notion that BHs evolve jointly with their host galaxies, mediated perhaps through 
feedback from active galactic nuclei (AGNs; e.g., Di~Matteo et al. 2005). Many issues affect 
our detailed understanding of the relationship between BHs and their hosts, including (1) 
the evolution of BH mass, accretion rate, and spin, (2) the assembly and evolution of galaxies, and
(3) the mutual interaction between BHs and galaxies. A simple comparison of the mass density of BHs 
in local galaxies with the integrated energy density of quasars (So\l tan 1982) leads to the conclusion 
that radiatively efficient accretion, with an average efficiency of $\eta \approx 0.1$, is the main 
driver of BH growth (Small \& Blandford 1992; Chokshi \& Turner 1992; Yu \& Tremaine 2002; Marconi 
et al. 2004).  If the angular momentum of the accreting material is aligned with the spin of the BH 
and remains constant, we expect the BH to spin up to its maximum rate after accreting one-third of 
its mass, and attain $\eta \approx 0.4$ (Thorne 1974), in apparent conflict with the efficiency 
deduced from So\l tan's argument.  This inconsistency can be plausibly resolved if BHs are fed by 
episodic, random accretion (King \& Pringle 2006), which has a tendency to spin down the BH over 
time (King et al. 2008). If this, indeed, is the main mechanism for BH growth, we predict that the 
radiative efficiency of quasars should decrease from high to low redshift.  The main goal of 
this paper is to test this prediction observationally.  By comparing the cosmic history of BH 
accretion with the better-established cosmic history of the star formation rate, we can 
further constrain the purported joint evolution of central BHs and their host galaxies
(e.g., Silk \& Rees 1998; Croton et al. 2006; Wang et al. 2007). 

In this Letter, we set up a formalism to describe the dependence of $\eta$ on the luminosity density, 
growth rate, and duty cycle of quasars.  We apply it to survey data and show that $\eta$ evolves 
strongly with redshift, which we interpret as a signature of BH spin-down as a consequence of episodic, 
random accretion.  Our calculations assume a standard cosmology with $H_0=71~{\rm km~s^{-1}~Mpc^{-1}}$, 
$\Omega_{m}=0.3$, and $\Omega_{\Lambda}=0.7$.

\section{The $\eta-$equation}
Following So\l tan's argument, we assume that the growth of BHs comes mainly from accretion during
their active (AGN) phases. From its definition (Marconi et al. 2004), the radiative efficiency is 
\begin{equation}
\eta=\frac{\Delta \epsilon}{\Delta \epsilon+\Delta \rhobh c^2},
\end{equation}
where $\Delta \epsilon$ and $\Delta \rhobh$ are the cumulative increases of the energy  
and the mass density of BHs from $z+\Delta z$ to $z$, and $c$ is the speed of light. 
We can define the duty cycle as the ratio of the number density of quasars ($N_{\rm Q}$)
to that of active and inactive galaxies ($N_{\rm G}$),
\begin{equation}
\delta(z)=\frac{N_{\rm Q}(z)}{N_{\rm Q}(z)+N_{\rm G}(z)},
\end{equation}
which can be equivalently expressed as (Wang et al. 2006, 2008)
\begin{equation}
\delta(z)=\frac{\rhobh^{\rm qso}(z)}{\rho_{\rm acc}(z)}, 
\end{equation}
where $\rhobh^{\rm qso}(z)$ is the mass density of quasar BHs and
$\rho_{\rm acc}(z)$ is the mass density of all gas accreted onto BHs until $z$.
The mass density of quasar BHs can be obtained from 
$\rhobh^{\rm qso}(z)=\int_{\mbh^*}\mbh\Phi(z,\mbh)\rmd \mbh$, 
where $\Phi(z,\mbh)$ is the mass function of BHs with the break given by $\mbh^*$. Since quasars 
can increase their masses only via accretion (i.e. $\Delta \rhobh=\Delta \rho_{\rm acc}$), we 
have, using the equivalent formulation of the duty cycle, the cumulative increase of the mass 
density 
\begin{equation}
\Delta \rhobh=\Delta \left(\delta^{-1} \rhobh^{\rm qso}\right).
\end{equation}
The cumulative energy density can be obtained through 
\begin{equation}
\Delta \epsilon=\dot{U}(z)\left(\frac{\rmd t}{\rmd z}\right) \Delta z,
\end{equation}
where $t$ is cosmic time, $\dot{U}(z)=\int_{L_{\rm Q}^*}^{\infty}\lbol\Psi_{\rm Q}(z,\lbol) \rmd \lbol$ 
is the bolometric luminosity density, $\Psi_{\rm Q}(z,\lbol)$ is the bolometric luminosity function 
(LF) of quasars (Hopkins et al. 2007), and the lower limit $L_{\rm Q}^*$, given by the break in the LF,
is determined from quasar surveys. We then have the $\eta-$equation:
\begin{equation}
\eta^{-1}=1+\frac{c^2}{\dot{U}}\left(\frac{\rmd t}{\rmd z}\right)^{-1}
          \frac{\rmd}{\rmd z} \left(\frac{\rhobh^{\rm qso}}{\delta}\right).
\end{equation}

Equation (6) is striking in that it describes $\eta$ in terms of $\rhobh^{\rm qso}/\delta$, which combines 
the growth and the episodic cycles of BHs with the evolution of their host galaxies. This equation has the 
advantage of only depending on observables: (1) $\rhobh^{\rm qso}(z)$, which can be reasonably estimated 
from empirical relations calibrated against reverberation mapping; (2) $\delta(z)$, which in principle 
can be estimated from deep galaxy surveys; and (3) $\dot{U}(z)$, which can be directly obtained from quasar 
LFs. Furthermore, the equation directly gives $\eta$ at any redshift, whereas the usual So{\l}tan's argument 
requires a comparison of the accreted mass density of BHs with the local mass density of BHs to get the 
averaged radiative efficiency over time (So{\l}tan 1982; Yu \& Tremaine 2002; Marconi et al. 2004). These 
features allow us to deduce the radiative efficiency as a function of redshift from observables. 

\section{Evolution of the Radiative Efficiency}
Equation (6) can be applied to existing survey data. We use the duty cycle expressed by $N_{\rm Q}(z)$ and 
$N_{\rm G}(z)$, where $N_{\rm Q}(z)=\int_{L_{\rm Q}^*}\Psi_{\rm Q}(z,L)\rmd L$ and 
$N_{\rm G}(z)=\int_{M_R^*}\Psi_{\rm G}(z,M_R)\rmd M_R$ are the number densities of quasars and galaxies
from their luminosity functions $\Psi_{\rm Q}$ and $\Psi_{\rm G}$, respectively. It should be noted that
$\Psi_{\rm G}$ is wavelength-dependent (Wolf et al. 2003), being less sensitive to 
stellar population changes at longer wavelengths (Ilbert et al. 2005). We calculate $N_{\rm G}(z)$ from 
the rest-frame $R-$band LFs of galaxies up to $z\approx 2$ derived from the deep 
{\em HST}\ study of Dahlen et al. (2005) and the VIMOS-VLT Deep Survey (VVDS) of Ilbert et al. (2005).
Both samples have similar selection criteria and photometric redshift estimates from $z\approx 0$ to 2. 
$N_{\rm Q}(z)$ is calculated from Bongiorno et al. (2007), who supplement the Sloan Digital Sky Survey 
(SDSS) quasar sample of Richards et al. (2006) with VVDS data for low-luminosity AGNs.
We calculate $\rhobh^{\rm qso}$ with BH masses estimated from the ``virial'' method, as given in 
Vestergaard et al. (2008), but use the corrected value for $z=0.49$ and the new 
value for 
$z=0.17$ 
in Kelly et al. (2009). The samples of quasars and galaxies used in these LFs overlap in at least 
some survey fields, minimizing potential observational biases. We relate the lower limits of the galaxy 
LF ($M_R^*$) and the BH mass function ($\mbh^*$) through the BH-bulge mass relation of McLure \& Dunlop 
(2002), which we assume to be invariant with redshift (cf. Ho 2007).  The completeness of the BH mass 
function depends on both the flux limit of SDSS and the widths of the broad emission lines.
The adopted BH mass function of Vestergaard et al. (2008), complete only for $\mbh$ \gax\ $10^{8.2}\sunm$,
yields $M_R^*=-22.4$ mag, which corresponds to $L_{\rm Q}^*=10^{45.5}$ erg~s$^{-1}$ if bright quasars 
have an average Eddington ratio of $\sim 0.2$ (Shen et al. 2008). 

\figurenum{1}
\centerline{\psfig{figure=fig1.ps,angle=270,width=8.0cm}}
\figcaption{\footnotesize Evolution of ({\em a}) the luminosity density, ({\em b}) 
quasar BH mass density, and ({\em c}) duty cycle. In calculating $\dot{U}$, 
the contribution from the infrared band, which represents reprocessed 
emission, has been subtracted. The solid lines represent the best 
linear-squares fits used for calculating $\eta$. In panel {\em b}, the red 
square ($\rhobh^{\rm qso}=10^{1.3}\sunm{\rm Mpc^{-3}}$) is the mass density of 
local active BHs (Greene \& Ho 2007). The final results ($\eta$), within the errors, 
do not depend on the particular analytical fits of $\rho_{\bullet}^{\rm qso}(z)$ and 
$\delta(z)$. Our calculations assume the following Poisson statistical uncertainties: 
$\Delta \Psi_{\rm Q}/\Psi_{\rm Q}=0.1$, $\Delta \Psi_{\rm G}/\Psi_{\rm G}=0.1$, and 
$\Delta \rhobh^{\rm qso}/\rhobh^{\rm qso}=0.1$, except for the red point in ({\em b}), 
for which we adopt $\Delta \rhobh^{\rm qso}/\rhobh^{\rm qso}=0.2$.
}
\label{fig1}
\vglue 0.3cm

\begin{center}{Table 1 Duty Cycles}
{\footnotesize
\begin{tabular}{rrrr}\hline\hline
$a_i$  &$M_{R}^*=-22.3$&$M_{R}^*=-22.4$&$M_{R}^*=-22.5$\\ \hline
$a_0$  &$-$4.10~~~~~~ &$-$4.01~~~~~~ &$-$3.92~~~~~~\\
$a_1$  &   3.34~~~~~~ &   3.32~~~~~~ &   3.29~~~~~~\\
$a_2$  &$-$1.60~~~~~~ &$-$1.59~~~~~~ &$-$1.57~~~~~~\\
$a_3$  &   0.26~~~~~~ &   0.26~~~~~~ &   0.25~~~~~~\\ \hline
\end{tabular}}
\end{center}

\figurenum{2}
\begin{figure*}[t]
\centerline{\includegraphics[angle=-90,width=17.50cm]{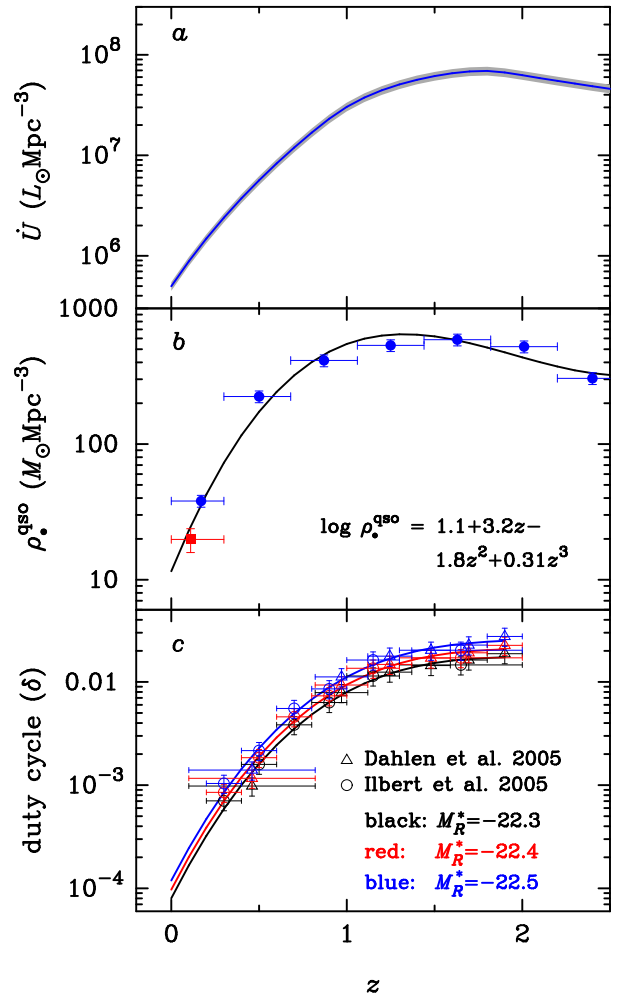}}
\figcaption{\footnotesize Evolution of ({\em a}) the luminosity density, ({\em b}) 
quasar BH mass density, and ({\em c}) duty cycle. In calculating $\dot{U}$, 
the contribution from the infrared band, which represents reprocessed 
emission, has been subtracted. The solid lines represent the best 
linear-squares fits used for calculating $\eta$. In panel {\em b}, the red 
square ($\rhobh^{\rm qso}=10^{1.3}\sunm{\rm Mpc^{-3}}$) is the mass density of 
local active BHs (Greene \& Ho 2007). The final results ($\eta$), within the errors, 
do not depend on the particular analytical fits of $\rho_{\bullet}^{\rm qso}(z)$ and 
$\delta(z)$. Our calculations assume the following Poisson statistical uncertainties: 
$\Delta \Psi_{\rm Q}/\Psi_{\rm Q}=0.1$, $\Delta \Psi_{\rm G}/\Psi_{\rm G}=0.1$, and 
$\Delta \rhobh^{\rm qso}/\rhobh^{\rm qso}=0.1$, except for the red point in ({\em b}), 
for which we adopt $\Delta \rhobh^{\rm qso}/\rhobh^{\rm qso}=0.2$.
}
\label{fig1}
\end{figure*}

Figure~1 shows the results for $\dot{U}(z)$, $\rhobh^{\rm qso}(z)$ and $\delta(z)$.  Panel ({\it b}) 
gives the best fit for $\rhobh^{\rm qso}(z)$. 
Since the duty cycle $\delta(z)$ depends on $M_R^*$, we plot the results for three choices of 
$M_R^*$ (the dependence on $L_{\rm Q}^*$ is absorbed into $M_R^*$). We fit the variation of 
$\delta$ with $z$ using a polynomial $\log \delta=\sum_{i=0}^{3}a_iz^i$; 
the coefficients 
$a_i$ are given in Table 1. We find that $\delta(z)$ is very similar for the samples of Dahlen 
et al. (2005) and Ilbert et al. (2005), indicating the robustness of the present results.
The large uncertainties in the photometric redshifts for low-$z$ galaxies 
preclude us from deriving $\delta$ directly for $z<0.3$; in this regime, 
$\delta$ is extrapolated from the analytical fits at higher redshift.  We 
hope to remedy this situation with future surveys.

Our main results are summarized in Figure~2{\em a}.  We find (1) $\eta\approx 0.3$ at $z\approx 2$, 
and (2) a strong cosmological evolution, with $\eta$ decreasing to $\sim$0.03 by $z\approx 0$. The 
low value of $\eta$ in the local Universe agrees with the results from the continuity equation of the 
BH number density (Merloni \& Heinz 2008). The high values of $\eta$ at large $z$ are consistent with 
the estimate from the cosmic X-ray background (Elvis et al. 2002), which is mainly 
contributed by AGNs at $z\approx 1-2$. At this epoch, the spin angular momentum of BHs may 
originate mainly from the orbital momentum in major BH coalescences (i.e. BH binaries 
with mass ratio $\sim 1$; those with mass ratio $< 1$ are referred to as minor coalescences)
(Hughes \& Blandford 2003). 

The strong evolution in $\eta$ provides important constraints on the accretion history of BHs. There 
are two ways for BHs to grow (Berti \& Volonteri 2008): (1) major and minor BH coalescences 
and (2) accretion of gas. While major coalescences do contribute some angular momentum to BHs (Hughes 
\& Blandford 2003), accretion is the main contributor to their masses (King et al. 2008). 
An initially non-rotating BH rotates very fast after accreting about 
one-third of its mass with the same sign of angular momentum (Thorne 1974), 
and its radiative efficiency approaches the maximum value of
$\eta_{\rm max}=0.42$. However, the net angular momentum contributed to accreting 
BHs depends on whether the accretion is prograde or retrograde, as described below.

\section{BH Growth by Random Accretion}

Supposing that accretion occurs randomly and episodically in quasars, then, at episode $i$ with angular 
momentum $\Delta \vec{\ell}_i$ and mass $\Delta m_i$, we have (as in a random walk), after 
$n$ episodes, the net angular momentum $\Delta\vec{L}_{\bullet}=\sum_{i=1}^n \Delta \vec{\ell}_i$ and
$\Delta L_{\bullet}^2=\sum_{i=1}^n \Delta \ell_i^2=n\Delta \ell^2$,
whereas the growth in mass $\Delta M_{\bullet}=\sum_{i=1}^n\Delta m_i=n\Delta m$.
The specific angular momentum of an accreting BH,
\begin{equation}
a\propto \frac{\sqrt{n}\Delta \ell}{n\Delta m}\propto n^{-1/2},
\end{equation}
decreases with the number of accretion events.  Consistent with the numerical simulations of
King et al. (2008), we expect $a$ to decrease with decreasing $z$.  The evolution of 
$\eta$ offers strong observational support for this picture.
The episode number $n$ can be estimated by 
\begin{equation}
n=\frac{\Delta M_{\bullet}}{\Delta m}=\delta(z)\frac{t_{\rm C}(z)}{\Delta t_0}, 
\end{equation}
where $\Delta t_0$ is the episode lifetime and $t_{\rm C}(z)$ is the age of the Universe at redshift 
$z$. For $z\approx 2$,
we have $\delta\approx 10^{-1.5}$ (Fig.~1{\it c}), $t_{\rm C}=2.0~$Gyr, and $\Delta t_0=1.0~$Myr as
estimated from the proximity effect of $z\approx 2$ quasars (Kirkman \& Tytler 2008), and hence
$n\approx 10^2$.  The parameter
$a$ decreases by a factor $\sim 10$ from $z\approx 2$ to $z\approx 0$. On the other hand,  
random accretion may occur through disks whose masses are limited by self-gravity (King et al. 2008)
to $\Delta M_{\rm sg}\approx \left(H/R\right)\mbh$, where $H$ is the height of the disk at radius $R$. 
We have 
\begin{equation}
n=f\left(\frac{H}{R}\right)^{-1}\approx 10^2, 
\end{equation}
where $H/R\approx 10^{-2}$ and $f=\Delta \mbh/\mbh\approx 1$ is the factor by which the BH mass has been 
increased due to accretion. The rough consistency between these two estimates of $n$ suggests that the
random accretion events originate from a self-gravitating disk. Additionally, the gas accreted onto BHs 
is only a tiny fraction of the gas content of the entire galaxy 
($\sim 10^{-3}$), so the net angular momentum from accretion should be independent of the
total angular momentum. The evolution of the radiative efficiency found here does not favor 
accretion with consistently prograde angular momentum, as would occur if the typical accretion 
event has much larger aligned angular momentum than the BH (Volonteri et al. 2007; Berti \&
Volonteri 2008).  Rather, this study strongly favors a picture in which most of the mass 
in BHs was gained through random accretion.

Accretion rates of $\sim 10-100 \sunm$~yr$^{-1}$ are needed to generate 
the prodigious luminosities seen in quasars.  Such large amounts of fuel 
cannot be sustained by internal sources (e.g., stellar mass loss); they almost 
certainly require external triggers, in the form of major gas-rich galaxy 
mergers and interactions.  How do major mergers lead to episodic, random 
accretion?  Numerical simulations and direct observations (see Schweizer 1998 
for review) show that gas-rich mergers drive large quantities of 
gas to the circumnuclear region of the merger remnant, where a large portion 
of the fuel gets consumed in a starburst.  The mechanisms by which gas gets 
further funneled to smaller scales relevant for feeding AGNs are unclear 
(see Wada 2004 for a review).  But two things are certain:  the gas in 
the circumnuclear region is clumpy, and in order for the gas to accrete,
it must get rid of most of its angular momentum.  This can be accomplished by 
clump-clump collisions, especially in a turbulent environment.  Such a 
process is inherently random, and the material that manages to fall in 
carries little or no memory of its original angular momentum.  Thus, despite 
the fact that the original source of gas derives from a major merger, the 
final, individual accretion events are essentially random, small fragments.

\section{Summary and Concluding Remarks}

We derive a simple formalism to link the radiative efficiency parameter $\eta$ 
with the luminosity density, BH mass density, and duty cycle of quasars.  Applying 
this formalism to existing survey data of quasars and galaxies, we
show that $\eta$ evolves strongly with redshift, 
with values of $\sim 0.3$ at $z \approx 2$ and $\sim 0.03$ at $z \approx 0$.
A straightforward interpretation of this trend is that the BHs in the quasar 
population begin with high spins at early times, but they gradually spin 
down to the present day.  We suggest that accretion events onto BHs are 
inherently random and episodic, and that this mode of accretion naturally
reduces their spin.

We end with a couple of corollary remarks.

From the expression for $\eta(z)$, we can derive $\dot{\rho}_{\bullet}(z)=\dot{U}(z)/\eta(z)c^2$,
the cosmic history of BH accretion. Figure~2{\em b} shows that $\dot{\rho}_{\bullet}(z)$ 
closely traces $\dot{\rho}_{\rm SFR}(z)$, the cosmic history of the star formation rate density. 
In agreement with earlier studies (e.g., Richstone et al. 1998), this suggests that the 
history of BH growth generally follows the history of stellar mass build-up in galaxies, 
at least for redshifts up to $\sim 2$.

The physical mechanism responsible for the generation of strong radio jets in 
radio-loud AGNs remains largely a mystery.  BH spin is often implicated as a 
critical parameter (e.g., Sikora et al. 2007).  The principal result of 
this study is that the average spin of BHs in quasars decreases with decreasing 
redshift.  If a large spin is required to produce strong jets, we would 
naively expect the fraction of radio-loud quasars to rise toward higher redshift.  
This is the opposite of what is actually observed in the SDSS sample of quasars 
studied by Jiang et al. (2007).

\acknowledgements{V. Wild is thanked for useful discussions on galaxy luminosity functions. 
LCH acknowledges a helpful conversation with C. Gammie on the scenario of random accretion.
We appreciate the stimulating discussions among the members of the IHEP AGN
group. The research is supported by NSFC via NSFC-10325313 and 10733010, and
by CAS via 10521001, KJCX2-YW-T03, and the 973 project.}


\begin{thebibliography}{}
\bibitem[]{}Berti, E., \& Volonteri, M. 2008, \apj, 684, 822
\bibitem[]{}Bongiorno, A., et al. 2007, \aap, 472, 443
\bibitem[]{}Chokshi, A., \& Turner, E. 1992, \mnras, 259, 421
\bibitem[]{}Croton, D., et al. 2006, \mnras, 365, 11
\bibitem[]{}Dahlen, T., et al. 2005, \apj, 631, 126
\bibitem[]{}Di Matteo, T., Springel, V., \& Hernquist, L. 2005, Nature, 430, 604
\bibitem[]{}Elvis, M., Risaliti, G., \& Zamorani, G. 2002, \apj, 565, L75
\bibitem[]{}Ferrarese, L., \& Merritt, D. 2000, \apj, 539, L9
\bibitem[]{}Gebhardt, K., et al. 2000, \apj, 539, L13
\bibitem[]{}Greene, J. E., \& Ho, L. C. 2007, \apj, 667, 131
\bibitem[]{}Ho, L. C. 2007, \apj, 669, 821 
\bibitem[]{}Hopkins, A. M., \& Beacom, J. F. 2007, \apj, 651, 142
\bibitem[]{}Hopkins, P. F., Richards, G. T., \& Hernquist, L. 2007, \apj, 654, 731
\bibitem[]{}Hughes, S. C., \& Blandford, R. D. 2003, \apj, 585, L101
\bibitem[]{}Ilbert, O.,  et al. 2005, \aap, 439, 863
\bibitem[]{}Jiang, L., Fan, X., Ivezi\'c, Z., Richards, G. T., Schneider, D. P., Strauss, M. A., 
\& Kelly, B. C. 2007, \apj, 656, 680 
\bibitem[]{}Kelly, B., Vestergaard, M., \& Fan, X. 2009, \apj, 692, 1388
\bibitem[]{}King, A. R., \& Pringle, J. E. 2006, \mnras, 373, L90
\bibitem[]{}King, A. R., Pringle, J. E., \& Hofmann, J. A. 2008, \mnras, 385, 1621
\bibitem[]{}Kirkman, D., \& Tytler, D. 2008, \mnras, 391, 1457
\bibitem[]{}Kormendy, J., \& Richstone, D. 1995, \araa, 33, 581
\bibitem[]{}Magorrian, J., et al. 1998, \aj, 115, 2285
\bibitem[]{}Marconi, A., \& Hunt, L. K. 2003, \apj, 589, L21
\bibitem[]{}Marconi, A., Risaliti, G., Gilli, R., Hunt, L. K., Maiolino, R., 
\& Salvati, M. 2004, \mnras, 351, 169
\bibitem[]{}McLure, R. J., \& Dunlop, J. S. 2002, \mnras, 331, 795
\bibitem[]{}Merloni, A., \& Heinz, S. 2008, \mnras, 388, 1011
\bibitem[]{}Richards, G. T., et al. 2006, \aj, 131, 2766
\bibitem[]{}Richstone, D., et al. 1998, \nat, 395, 14
\bibitem[]{}Schweizer, F. 1998, in Saas-Fee Advanced Course 26, Galaxies: Interactions and Induced 
Star Formation,  ed. R. C. Kennicutt Jr., et al. (Berlin: Springer-Verlag), 105
\bibitem[]{}Shen, Y., Greene, J. E., Strauss, M. A., Richards, G. T., \& 
Schneider, D. P. 2008, \apj, 689, 160
\bibitem[]{}Sikora, M., Stawarz, L., \& Lasota, J.-P. 2007, \apj, 658, 815
\bibitem[]{}Silk, J., \& Rees, M. J. 1998, \aap, 331, L1
\bibitem[]{}Small, T.~A., \& Blandford, R.~D. 1992, \mnras, 259, 725
\bibitem[]{}So{\l t}an, A. 1982, \mnras, 200, 115
\bibitem[]{}Thorne, K. S. 1974, \apj, 191, 507
\bibitem[]{}Tremaine, S., et al. 2002, \apj, 574, 740
\bibitem[]{}Vestergaard, M., Fan, X., Tremonti, C. A., Osmer, P. S., \& 
Richards, G. T. 2008, \apj, 674, L1
\bibitem[]{}Volonteri, M., Sikora, M., \& Lasota, J.-P. 2007, \apj, 667, 704
\bibitem[]{}Wada, K. 2004, in Carnegie Observatories Astrophysics Series, Vol. 
1: Coevolution of Black Holes and Galaxies, ed. L. C. Ho (Cambridge: 
Cambridge Univ. Press), 186
\bibitem[]{}Wang, J.-M., Chen, Y.-M., Yan, C. S. \& Hu, C. 2008, \apj, 673, L9
\bibitem[]{}Wang, J.-M., Chen, Y.-M., Yan, C. S., Hu, C., \& Bian, W.-H. 
2007, \apj, 661, L43 
\bibitem[]{}Wang, J.-M., Chen, Y.-M., \& Zhang, F. 2006, \apj, 647, L17
\bibitem[]{}Wolf, C., et al. 2003, \aap, 401, 73
\bibitem[]{}Yu, Q., \& Tremaine, S. 2002, \mnras, 335, 965
\end{thebibliography}
\end{document}